\documentclass[aps,prb,twocolumn,superscriptaddress,showpacs]{revtex4-2}
\usepackage[utf8]{inputenc}
\usepackage{longtable}
\usepackage{amsmath}%
\usepackage{amsfonts}%
\usepackage{amssymb}%
\usepackage{graphicx}
\usepackage[hidelinks, breaklinks=true]{hyperref}
\usepackage{CJKutf8} 
\usepackage{ulem} 
\usepackage[product-units=power, separate-uncertainty=true, multi-part-units=single]{siunitx} 
\usepackage{xspace} 
\usepackage{xcolor}

\newcommand{\myref}[7]{\href{http://dx.doi.org/#7}{#1, #2, #3 \textbf{#4}, #5 (#6).}} 
\newcommand{\Gnull}{G-$R0^\circ$\xspace}

\newcommand{\Gdrei}{G-$R30^\circ$\xspace}

\newcommand{\sixsqthree}{$\left(6\sqrt{3}\times6\sqrt{3}\right)R\ang{30}$\xspace}

\newcommand{\ZLGdrei}{ZLG-$R30^\circ$\xspace}
 
\newcommand{\etal}{\textit{et al.}\xspace}

\newcommand{\pgi}{Peter Grünberg Institut (PGI-3), Forschungszentrum Jülich, 52425 Jülich, Germany}
\newcommand{\jara}{Jülich Aachen Research Alliance (JARA), Fundamentals of Future Information Technology, 52425 Jülich, Germany}
\newcommand{\rwth}{Experimentalphysik IV A, RWTH Aachen University, 52074 Aachen, Germany}

\begin{document}

\title{
Epitaxial growth of mono- and (twisted) multilayer graphene on SiC(0001)
}

\author{Hao~Yin (\begin{CJK*}{UTF8}{bsmi}尹~浩\end{CJK*})}
\affiliation{\pgi} \affiliation{\jara} \affiliation{\rwth}
\author{Mark~Hutter}      
\affiliation{\pgi} \affiliation{\jara} \affiliation{\rwth}
\author{Christian~Wagner}
\affiliation{\pgi} \affiliation{\jara}
\author{F.~Stefan~Tautz} 
\affiliation{\pgi} \affiliation{\jara} \affiliation{\rwth}
\author{François~C.~Bocquet} 
\affiliation{\pgi} \affiliation{\jara}  
\author{Christian~Kumpf} \email{c.kumpf@fz-juelich.de}
\affiliation{\pgi} \affiliation{\jara} \affiliation{\rwth}
\date{\today}

\begin{abstract}
To take full advantage of twisted bilayers of graphene or other two-dimensional materials, it is essential to precisely control the twist angle between the stacked layers, as this parameter determines the properties of the heterostructure. 
In this context, a growth routine using borazine as a surfactant molecule on SiC(0001) surfaces has been reported, leading to the formation of high-quality epitaxial graphene layers that are unconventionally oriented, i.e., aligned with the substrate lattice (\Gnull) [Bocquet et al. Phys. Rev. Lett. 125, 106102 (2020)].
Since the \Gnull layer sits on a buffer layer, also known as zeroth-layer graphene (ZLG), which is rotated $30^\circ$ with respect to the SiC substrate and still covalently bonded to it, decoupling the \ZLGdrei from the substrate can lead to high-quality twisted bilayer graphene (tBLG). Here we report the decoupling of \ZLGdrei by increasing the temperature during annealing in a borazine atmosphere. While this converts \ZLGdrei to \Gdrei and thus produces tBLG, the growth process at elevated temperature is no longer self-limiting, so that the surface is covered by a patchwork of graphene multilayers of different thicknesses. We find a 20\% coverage of tBLG on ZLG, while on the rest of the surface tBLG sits on one or more additional graphene layers. In order to achieve complete coverage with tBLG only, alternative ways of decoupling the ZLG, e.g. by intercalation with suitable atoms, may be advantageous.    
 \end{abstract}

\maketitle


\section{\label{sec:level1}Introduction}

When Geim and Novoselov published their groundbreaking work on exfoliated graphene sheets in 2004 \cite{Novoselov2004}, they opened up a new research field that has been growing ever since. Many novel and exotic properties have been theoretically predicted and experimentally observed in graphene and other two-dimensional (2D) van-der-Waals (vdW) materials. Maybe the most notable aspect is the fact that electronic properties of these materials can be tuned by proximity effects without essentially changing their geometric structure. One example is the concept of twistronics, i.e., the formation of twisted 2D bi- and multilayers forming specific Moiré superlattices \cite{Lopes2007, Shallcross2010, Bistritzer2010, Bistritzer2011, Cao2016, Kim2017}. The decisive parameter affecting the properties of these systems is the twist angle between the 2D layers. For twisted bilayer graphene (tBLG), intriguing properties like unconventional intrinsic superconductivity, a tunable band gap, strong interlayer coupling, topological states, the quantum Hall effect, etc., have been experimentally reported or theoretically proposed \cite{Kim2017, Cao2018, Yao2018, Lu2019, Rademaker2020, Chen2020, Ledwith2020, Repellin2020}, and different ways to grow such bilayers have been described \cite{Sutter2009, Sun2021, Yan2014}.
For small twist angles in graphene, the interlayer coupling decreases with increasing twist. In a certain intermediate range of angles, the bilayer film behaves more like independent monolayers \cite{Shallcross2010, Bistritzer2010}. However, at the highest possible twist angle, i.e., $30^{\circ}$, a strong interaction emerges again and the tBLG sample exhibits intriguing phenomenona, such as a mirrored Dirac cone, nearly flat bands, and dodecagonal quasicrystallinity \cite{Yao2018, Ahn2018, Moon2019}. 

$30^{\circ}$-rotated tBLG samples demand extremely precise orientation accuracy, since even a tiny deviation can destroy their long-range quasi-crystalline features \cite{Moon2019}. Precise angle control and reproducibility present challenges when employing the conventional micro-mechanical exfoliation method (``tear and stack'') to fabricate samples. Hence, alternative ways to realize $30^{\circ}$ rotated tBLG with high quality and uniform orientation have been tried, e.g., carbon segregation from a hot Pt(111) bulk crystal \cite{Yao2018}, or chemical vapor deposition (CVD). The latter results in epitaxially grown graphene on metal surfaces like, e.g., Cu foils \cite{Pezzini2020}, Cu(111) single crystals \cite{Deng2020, Cho2020} or Ni-Cu gradient alloys \cite{Takesaki2016, Gao2018}. However, in all cases, tBLG forms islands and does not cover the whole surface uniformly.

Growing epitaxial graphene on bulk SiC is a very promising approach that can be adapted to fabricate tBLG with orientational precision and homogeneously covering all of the surface. The flake size is then only limited by the size of the SiC terraces. The standard recipe is thermal decomposition of the uppermost SiC layers by annealing in UHV at temperatures above 1300$^{\circ}$C. Performed on a (0001) oriented crystal, this generates graphene layers that are rotated by $30^{\circ}$ with respect to the SiC substrate lattice. We designate such layers \Gdrei throughout this paper.  Graphene can be grown in multilayers using this technique, with the lowest layer not fully decoupled from the underlying SiC bulk, but still covalently bonded to the substrate. Hence, it is commonly called ``buffer layer'' or ``zeroth-layer graphene'' (ZLG). It exhibits a \sixsqthree-reconstructed superstructure \cite{Starke2009, Riedl2010}. 

Evidently, for achieving epitaxial tBLG on SiC, one has to find ways to grow graphene in different orientations. Ahn \etal \cite{Ahn2018} suggested a pathway to epitaxial $30^{\circ}$-rotated tBLG by first growing a boron nitride (BN) layer on 4H-SiC(0001). For this purpose, the SiC bulk crystal is annealed in borazine atmosphere at 1050$^{\circ}$C. The resulting BN layer is aligned with the bulk lattice (i.e., $0^{\circ}$-rotated), and carbonizes upon (post-growth) annealing at a higher temperature (1600$^{\circ}$C). Although the precise composition of this BN layer is unclear \cite{Lin2022}, it transforms into graphene while conserving its orientation, and hence the resulting graphene layer is oriented $0^{\circ}$ with respect to the substrate. Bocquet et al.\ \cite{Bocquet2020} demonstrated that such an unconventionally oriented graphene layer can also be grown in a one-step process. They annealed the 6H-SiC(0001) surface in borazine atmosphere at $>1300^{\circ}$C and obtained the \Gnull directly and in much higher quality. Apparently, in this case, the borazine molecules act as surfactants at the hot SiC surface and force the graphene into the unconventional $0^{\circ}$ orientation directly during the thermal SiC decomposition process.
In a comprehensive study based on spot profile analysis low-energy electron diffraction (SPA-LEED), angle resolved photoelectron spectroscopy (ARPES) and normal incidence X-ray standing wave technique (NIXSW), this single $0^{\circ}$ orientated graphene layer, designated \Gnull throughout this paper, has been demonstrated to exhibit both a high surface coverage and a high crystalline quality   \cite{Bocquet2020}. Here, we follow up on this work and extend it to multilayers, including tBLG. Our study is based on low energy electron microscopy (LEEM), a technique enabling \textit{in-situ} measurements during growth and/or thermal processing.


\section{Experimental Details}

\subsection{Sample Preparation}

N-doped 6H-SiC(0001) wafers, purchased from SITUS Technicals GmbH, Germany, were used for the preparation of the samples discussed here. In the first preparation step, the samples were thoroughly degassed in UHV at 880$^{\circ}$C for 30 min. 
Afterwards, while exposing the surface to an external supply of Si from an evaporator, the sample was annealed to 1050$^{\circ}$C, causing a $\left(\sqrt{3}\times\sqrt{3}\right)$-R$30^\circ$ reconstruction of the surface. Reducing the sample temperature to 880$^{\circ}$C while the Si exposure is maintained for 30 min, leads to the formation of a homogeneous, Si-rich $(3\times3)$ reconstructed surface. 
\Gnull was then obtained by annealing the sample at $T> 1300^{\circ}$C in a borazine atmosphere with a partial pressure of $1.5\times10^{-6}$ mbar for 30 min. The annealing temperature during this final step was the only control parameter used for the preparation of the samples discussed in this paper. Specifically, the sample denoted as \textit{low-T} in the following was annealed at $1330^{\circ}$C in borazine atmosphere, while for the \textit{high-T} sample the annealing temperature was $1380^{\circ}$C. The sample temperature was measured using a pyrometer with $\epsilon=0.825$.

\subsection{Methods}

Beside LEEM-based measurements, we also performed ARPES and LEED measurements. A LEED instrument (OCI model BDL800IR) mounted at the preparation chamber was used to check the sample surface quality after each preparation step.
The ARPES measurements were performed in an UHV analysis chamber directly connected to the preparation chamber. It is equipped with a Scienta R4000 hemispherical analyzer and a Scienta VUV-5k He-lamp. We used the He-I line ($21.2$~eV) for recording ARPES spectra at room temperature for all samples after final preparation. 

After preparation and performing the ARPES and LEED measurements, the samples were transported through air to the LEEM instrument and outgassed again at  $\sim 900^{\circ}$C in order to remove any contamination that occurred during transport. Operating the LEEM instrument in diffraction mode, we recorded LEED images to verify the good air-stability of the samples.

The LEEM study was performed in a spectroscopic photoemission and low-energy electron microscope (SPELEEM 3) of Elmitec GmbH. In addition to the diffraction mode, the instrument offers several different LEEM operating modes that are selected by positioning a contrast aperture in the electron beam path. We applied both the bright-field (BF) mode (whence the aperture is positioned such that only the (00) diffracted beam is used for imaging) and the dark-field (DF) mode (for which any other specific diffracted beam can be selected). The latter allows the identification of specific structures on the surface, as only those areas appear bright in the image that contribute to the selected beam. Areas with structures that scatter less (or not at all) into the selected beam appear darker (or black). 

A specific (spectroscopic) version of BF-LEEM imaging was applied throughout the present study: LEEM-$IV$ spectroscopy, in which the BF-LEEM image is recorded continuously while the interaction energy of the electrons (the so-called start energy) is varied from slightly negative values (mirror mode, the electrons are reflected before they reach the sample surface) up to values of several tens of eV. Usually, from these data, the $IV$ (intensity vs.\ start voltage) spectra of selected individual regions in the LEEM image are extracted. Here, however, we extracted these spectra pixel-by-pixel from the LEEM images, with the goal to determine how many graphene layers are present on the surface. For epitaxial graphene, it has been demonstrated that this number corresponds to the number of minima that occur in the $IV$ spectra in the energy range from $0$ to $7$~eV \cite{Hibino2008, Ohta2008, Luxmi2010, Mende2018, Riedl2009}. Hence, we restricted our measurements to this range. The oscillations in the reflected electron intensity result from interference effects between electron beams that are reflected from the various interfaces within the graphene stack. They can also be understood as the interlayer band of graphite that splits into discrete levels for few-layer graphene \cite{Hibino2008, Ohta2008, Luxmi2010, Mende2018, Riedl2009}. Note that this technique only detects the number of (quasi) free-standing graphene layers, i.e., electronically decoupled layers. ZLG cannot be detected in this way.

In our LEEM-$IV$ analysis, we utilized the spatial resolution of our instrument (about $8$~nm in uncorrected mode) to full extent: From the recorded three-dimensional data stack (intensity $I$ as a function of the direct space coordinates $x$ and $y$ and the start energy $E$, corresponding to the start voltage $V$), we extracted $I_{x,y}(E)$ curves pixel by pixel on the detector, i.e., for each pair $(x, y)$ in the image. This resulted in up to $\sim 260.000$ spectra, if full-size images with $512 \times 512$ pixels were used. After a minimum of data correction (essentially only drift correction), these spectra were then analyzed by an automatized algorithm based on the vector quantization method \textit{K-means} \cite{scikit-learn}, which sorts curves with similar shapes into a pre-selected number of clusters. According to their shape, in particular the number of minima in the relevant energy range, each of these clusters was then assigned to a specific number of graphene layers, and regions on the sample can be identified accordingly. Compared to similar methods in the literature \cite{Jong2019, Masia2022, Wolff2014}, our approach requires less pre-processing of the data. The result of the analysis is a false color image of equal size and resolution as the BF-LEEM images, with the colors encoding the number of graphene layers in the respective area. We note that this technique is model free, i.e., no reference curves are needed, and based on an unsupervised machine learning algorithm.


\section{Results and Discussion}

We discuss the structure and morphology of graphene layers on SiC(0001) after surfactant-mediated epitaxial growth in borazine atmosphere, with a strong focus on the influence of the temperature during borazine exposure. In particular, we report results from two samples that were prepared at similar temperatures ($1330^{\circ}$C and $1380^{\circ}$C), but nevertheless show fundamentally different morphologies. While a homogeneous \Gnull layer is formed on the ZLG at the lower temperature, at the higher temperature the ZLG decouples from the substrate. In principle, on the pathway to tBLG this is desired, since the ZLG is $30^\circ$-rotated with respect to the substrate, and hence forms a \Gdrei layer below the \Gnull layer. However, we find that the decoupling does not take place homogeneously, but results in an inhomogeneous surface with multilayer areas with different numbers of graphene layers. Stacks with up to five decoupled layers have been identified. 

In the following section, the high-T and low-T samples are compared according to their electronic and geometric structure, based on BF-LEEM, ARPES and LEED. Then, we separately report detailed LEEM studies of the morphology of either sample in two dedicated sections. Both of them include DF-LEEM and LEEM-$IV$ data and their analysis, and also introduce in detail our automatized algorithm for a cluster-analysis of LEEM-$IV$ data. 

\subsection{Homogeneity and Electronic Structure}
\label{sec:homogeneity}

\begin{figure*}
\includegraphics[width=\textwidth]{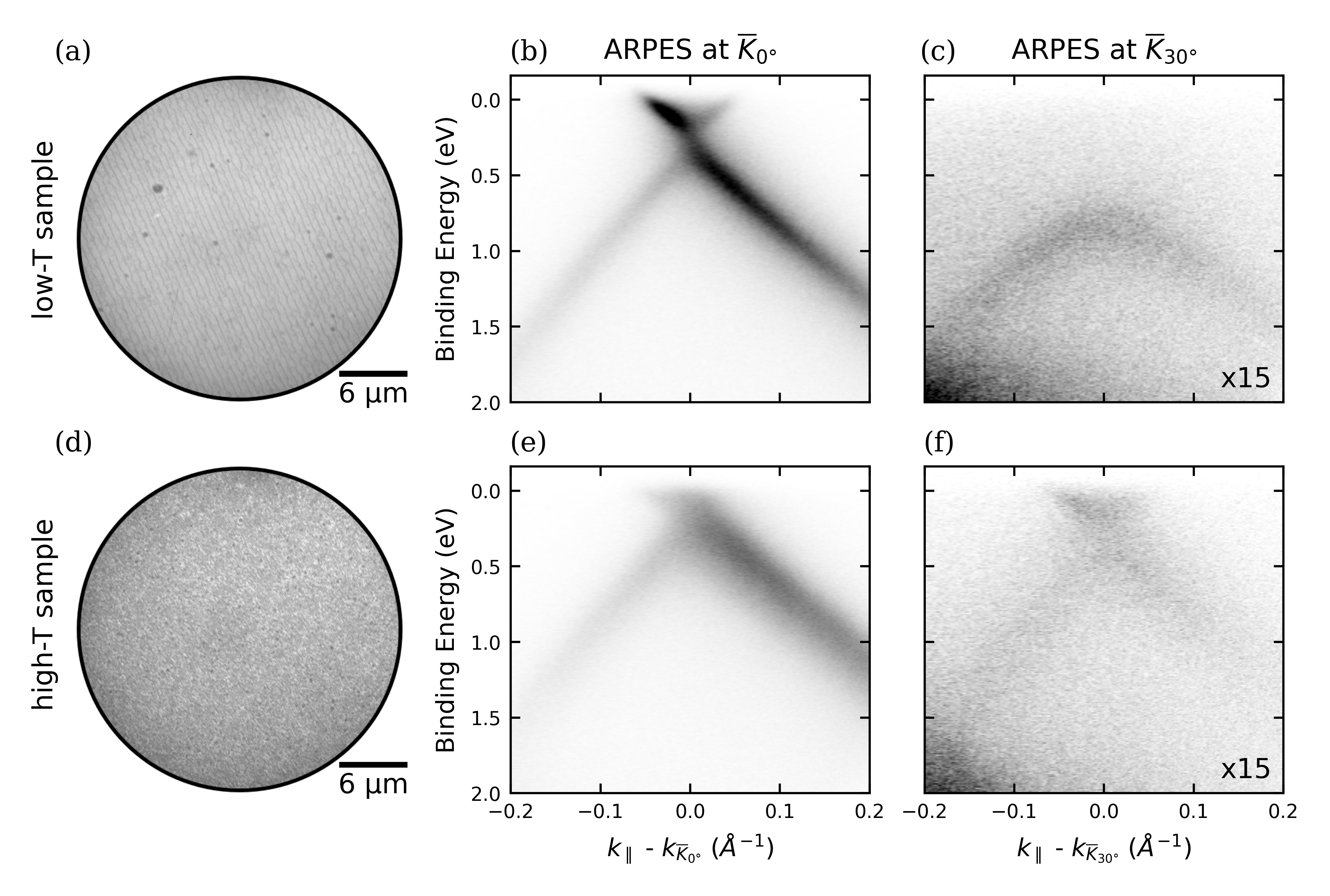}
\caption{\label{fig:1}(a) Bright field (BF) LEEM image of the low-T sample, taken at an electron energy (start energy) of $1.66$~eV with a field of view (FoV) of $30~\mu$m. (b-c) ARPES energy distribution maps (EDM) recorded along the $\bar{\Gamma}\bar{K}$~direction close to the $\bar{K}$ points of \Gnull and \Gdrei, respectively. (d-f) Same as panels (a-c), but  for the high-T sample.}
\end{figure*}

In Fig.\ \ref{fig:1} we present a comparison of the two samples based on ARPES data that has been measured directly after preparation (panels (b-c) and (e-f)), and BF-LEEM images recorded right after transfer of the sample into the microscope (panels (a) and (d)). The LEEM images show a field of view (FoV) of $30~\mu$m, but for both samples these images are representative for much larger areas ($>200~\mu$m) on the respective surface, showing similar morphologies. The surface of the low-T sample is very uniform, with parallel lines running across the FoV. They represent the step edges which separate the surface terraces of the SiC substrate. Only a few defects are visible as dark spots. In contrast, the surface of the high-T sample is more inhomogeneous, with changes between brighter and darker regions appearing on  a sub-$\mu$m length scale. The step edges are still faintly visible, but they are discontinuous and appear to be much denser than on the low-T sample surface. Furthermore, the density of contamination spots (dark) is higher than on  the low-T surface, but their size is smaller.

The electronic band structure in the ARPES band maps yields insight into the quality and the orientation of the graphene layers. Figure \ref{fig:1}(b) shows an energy distribution map (EDM) recorded on the low-T sample at the position in momentum space where the $\bar{K}$ point of \Gnull ($\bar{K}_{0^\circ}$) is expected. The characteristic Dirac cone of graphene is clearly visible, slightly shifted downwards (i.e., to higher binding energies), thus indicating a small $n$ doping. 
In contrast, the EDM in Fig.\ \ref{fig:1}(c), which was recorded where the $\bar{K}$ point of \Gdrei ($\bar{K}_{30^\circ}$) would be expected, i.e., $30^{\circ}$-rotated with respect to the EDM in Fig.\ \ref{fig:1}(b), shows no Dirac cone. The band visible at $>1$~eV binding energy is a band of \Gnull close to its M point. Note that this band is very weak, the data is scaled up by a factor of 15 compared to Fig.\ \ref{fig:1}(b).

On the high-T sample we observe a Dirac cone at $\bar{K}_{0^{\circ}}$, very similar to the one on the low-T sample (just slightly weaker), see Fig.\ \ref{fig:1}(e). Hence, also on this sample we find \Gnull, maybe a little less $n$ doped than in the case of the low-T sample. However, for the high-T sample we also very clearly see a Dirac cone at $\bar{K}_{30^{\circ}}$(Fig.\ \ref{fig:1}(f)), indicating a significant coverage of \Gdrei. The intensity of the Dirac cone at $\bar{K}_{30^{\circ}}$ is much weaker than that of the one at $\bar{K}_{0^{\circ}}$. This might indicate a smaller coverage of \Gdrei than of \Gnull, but we certainly also expect  the shielding of the photoemission from the \Gdrei by the \Gnull because of the specific growth mode: at higher temperatures the \ZLGdrei  underneath the surfactant-induced \Gnull layer decouples from the substrate and forms \Gdrei \textit{below} the \Gnull. 

We can thus conclude that the low-T sample consists of \Gnull, but no \Gdrei. We note that below the \Gnull, a \ZLGdrei layer -- covalently bound to the SiC substrate -- must be present according to the surfactant-mediated growth mechanism of \Gnull that has been studied in detail before \cite{Bocquet2020}. Since this mechanism is self-limiting, we can also conclude that there is only \textit{one} \Gnull layer present on the low-T sample, although the data in Fig.~\ref{fig:1}(b-c) do not allow to distinguish between one or more layers. Regarding the high-T sample, we conclude that \Gdrei is present below the \Gnull, although the data discussed up to this point do not allow us to quantify this coverage (less than one layer, a single layer or even more than one layer, all presumably on a covalently bonded \ZLGdrei layer). In this respect, the morphology of the high-T sample turns out to be revealing (see below). Indeed, the most notable difference between the two samples is their morphology. This is already obvious from the LEEM images shown in Fig.\ \ref{fig:1}(a) and (d), but DF-LEEM and LEEM-$IV$ studies reveal many more details, as we discuss in the following two sections.

\subsection{Morphology of the low-T \Gnull sample}

\begin{figure}
\includegraphics[width=\columnwidth]{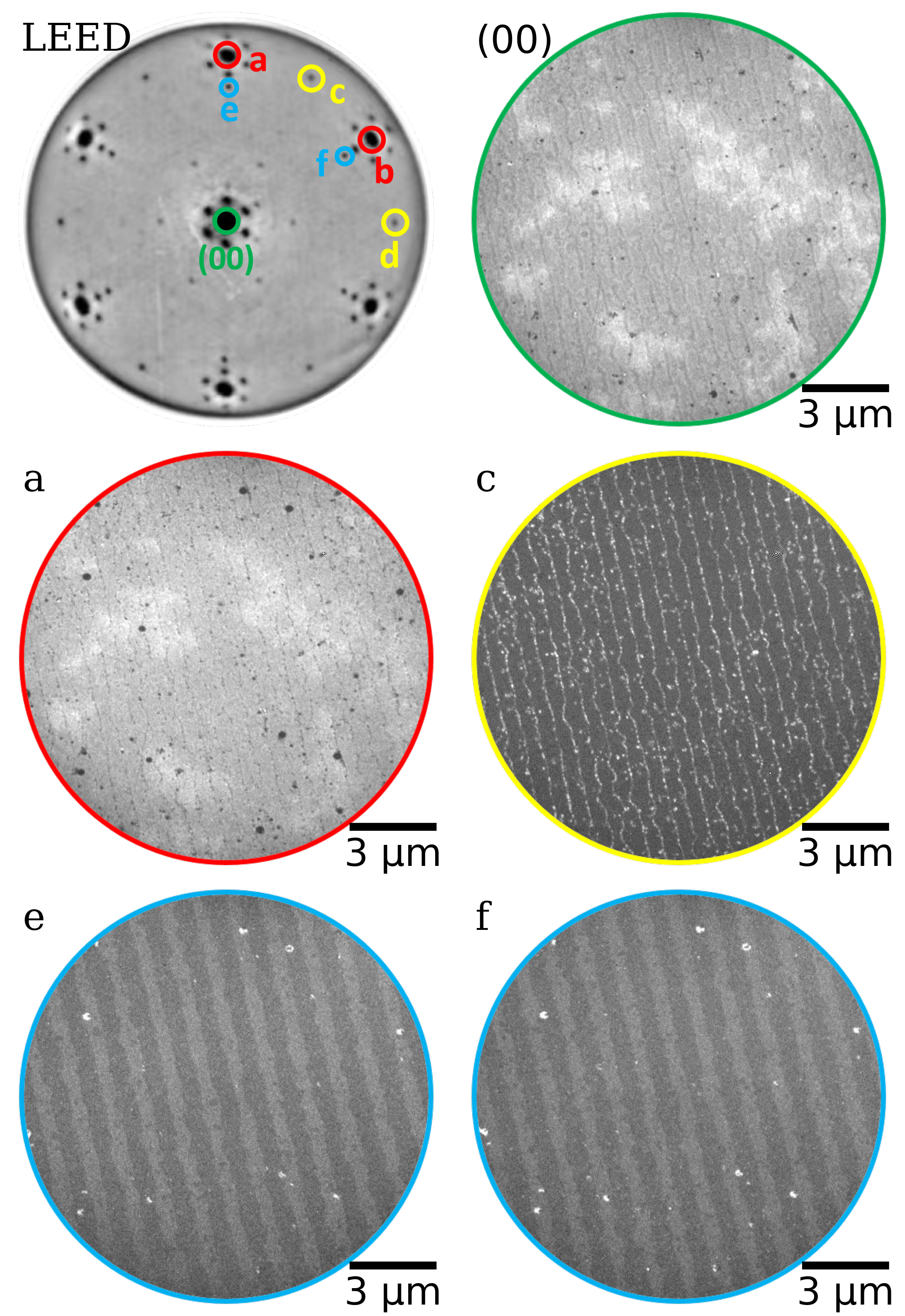}
\caption{\label{fig:2}LEED pattern of the low-T sample, and corresponding bright-field (BF) LEEM image, labeled  \textit{(00)}, and dark field (DF) LEEM images, labeled \textit{a}, \textit{c}, \textit{e}, and \textit{f}. The labels of the LEEM images correspond to the diffraction spots marked in the LEED pattern. All LEEM images were obtained at an electron energy (start energy) of $E=55$~eV.}
\end{figure}

The LEED pattern in the upper left of Fig.\ \ref{fig:2} contains diffraction spots originating from the SiC substrate (two of them marked by cyan colored circles and labeled \textit{e} and \textit{f}) and from \Gnull (red, labeled \textit{a} and \textit{b}). The spots from these two structures are only radially displaced with respect to each other, indicating that the corresponding lattices are orientationally aligned (both have $0^\circ$ orientation). A third set of reflections marked in yellow and labeled \textit{c} and \textit{d} is rotated by $30^\circ$ with respect to the substrate spots. Since the ARPES measurements unambiguously showed that there is no graphene layer with this orientation on this sample, these spots must originate from the \ZLGdrei. Also, other reflections visible in the LEED pattern stem from either the \sixsqthree reconstructed ZLG or from multiple scattering. Details of multiple scattering effects have been discussed in our previous work \cite{Bocquet2020}. 

Beside the LEED pattern, five LEEM images are displayed in Fig.\ \ref{fig:2}. The one labeled \textit{(00)} is a BF image (see Methods section above). 
In contrast to Fig.\ \ref{fig:1}(a), this BF image is recorded at a rather high electron energy ($E=55$~V), i.e., the same energy that was used for LEED and DF-LEEM. This image makes additional, dendritic features on the surface visible. They appear only at specific electron energies and are not caused by multilayer formation, as revealed by the LEEM-$IV$ measurements discussed below.

The other four images in Fig.\ \ref{fig:2} are DF-LEEM images obtained from the different diffraction spots labeled in the LEED pattern. Images based on spots \textit{a} or \textit{b} (as well as \textit{c} or \textit{d}) are pair-wise identical, as expected due to the six-fold symmetry of single-layer graphene. We hence show the DF-LEEM images for spots \textit{a} and \textit{c} only. Image \textit{a} is similar to the BF image \textit{(00)} (including the dendrites), while image \textit{c} appears homogeneously dark, apart from a few thin bright lines and tiny spots. This proves that the entire surface of  this sample is solely covered with \Gnull, while \ZLGdrei is only visible at small defects and possibly at the step edges. This observation is consistent with the expectation that the \ZLGdrei is located below the \Gnull layer. The similarity between the \textit{(00)} and \textit{a} images confirms that the dendritic contrast is not due to any other rotation of the graphene layer. 

The DF images recorded for the bulk spots (\textit{e} and \textit{f} in Fig.\ \ref{fig:2}) are not identical (in contrast to the pairs of images \textit{a}/\textit{b}  and \textit{c}/\textit{d} based on graphene spots), because SiC is only threefold symmetric. Instead, we observe bright and dark stripes in both images and a contrast inversion between \textit{e} and \textit{f}. This effect arises from the combination of the substrate symmetry with a certain type of substrate steps. Upon annealing, the steps on the 6H-SiC(0001) surface tend to bunch, forming steps that are three single layers high \cite{Nakajima2005, Pakdehi2020}. As a result, the neighboring terraces are mirror-symmetric in their surface termination, which together with the three-fold symmetry of the substrate produces the observed contrast inversion in the DF-LEEM images upon $60^\circ$ rotation. 

\begin{figure*}
\includegraphics[width=\textwidth]{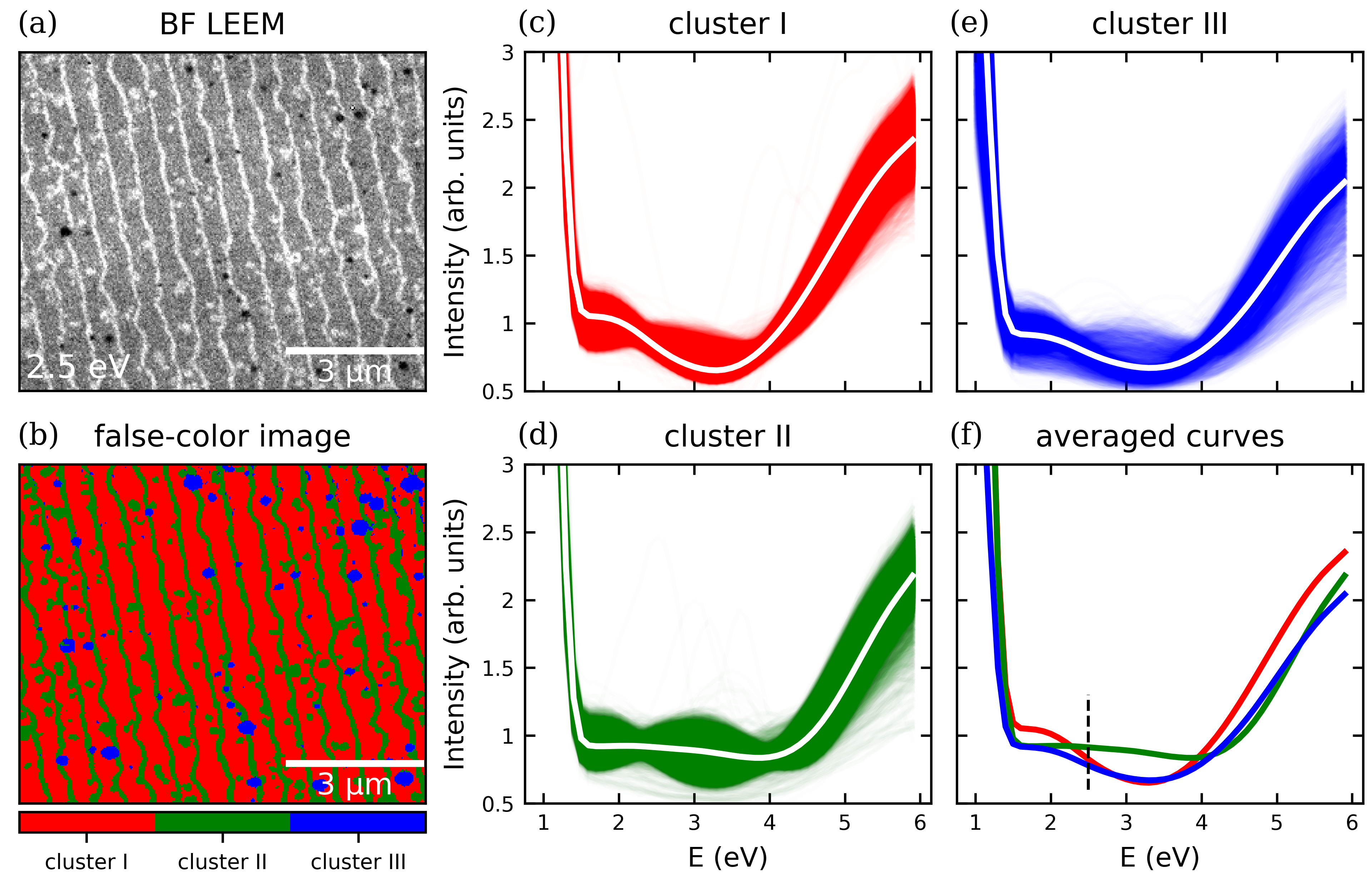}
\caption{\label{fig:3}(a) Bright-field (BF) LEEM image of the low-T sample, recorded at $E=2.5$~eV. LEEM-$IV$ data ($E=0$ to $5.9$~eV) have been recorded for this surface area  ($300\times 250$ pixels) pixel-by-pixel and analyzed using the \textit{K-means} algorithm  (with $K=3$) described in the text. (b) False-color image representing the result of clustering the LEEM-$IV$ data. (c-e) All LEEM-$IV$ spectra belonging to each of the three clusters. Individual spectra are plotted with a transparency of $0.01$. The white curves represent the average of all spectra in the respective cluster. (f) For clarity, the average curves from panels (c-e) are plotted in one diagram in their corresponding color. The dashed black line marks the energy at which the BF-LEEM image in panel (a) was recorded. }
\end{figure*}

For our LEEM-$IV$ analysis, we selected a $9.0 \times 7.5~\mu$m$^2$ section of the surface shown in Fig.\ \ref{fig:2}. BF-LEEM images of this section have been recorded for start energies from $E=0$~eV to $5.9$~eV, in steps of $0.1$ eV, i.e., at $60$ different energies.  
One representative image, recorded at $E=2.5$~eV, is shown in Fig.\ \ref{fig:3}(a). At this electron energy, the surface terraces look quite uniform. Step edges appear brighter than the terraces, and both bright and dark defects are visible.
From such a data stack $I(E,x,y)$, LEEM-$IV$ data are usually extracted by integrating the intensity within a subjectively chosen homogeneous part of the image and plotting this integrated intensity versus energy.  In this way, for each relevant area on the surface, a single representative spectrum is found. Whenever there are a limited number of clearly identifiable homogeneous areas on the sample, this approach yields unambiguous and relevant results, although it tends to systematically ignore regions close to the borders between different homogeneous areas. But for samples that do not have clearly defined homogeneous regions (such as the one in Fig.~\ref{fig:1}(d)), the selection of approximately homogeneous surface areas by a human is prone to introduce bias. We therefore applied a generic way of extracting and analyzing the $IV$ spectra that avoids any bias and allows an objective identification of regions on the surface in which the $IV$ spectra are similar. For this purpose, we first extracted the spectra pixel-by-pixel from the complete data stack, i.e., for each pixel we obtained one individual LEEM-$IV$ curve. In our case, for data recorded with a resolution of $300\times250$ pixels, this resulted in 75.000 $I(E)$ curves, each with $60$ data points (energies). 
Second, these spectra were sorted into classes by a classification algorithm. We tested various approaches, including human identification of unique features in the spectra (such as certain curvatures or the numbers and positions of minima and turning points) that were then used for the classification---with moderate success. We therefore settled for a model-free and fully automatized classification process. Specifically, we applied the clustering algorithm \textit{K-means} \cite{scikit-learn}, which is well established for grouping vectors into a given number of clusters $K$, the only free parameter of the algorithm. The algorithm considers each $IV$ spectrum as a vector in an $n$-dimensional Euclidean space, where $n$ is the number of data points (energies) of the $IV$ spectra, i.e., $n=60$ in our case. 
The algorithm then places all vectors ($IV$ spectra) into clusters so that each of them belongs to the cluster with the nearest mean. Technically, the sum of the squared Euclidean distances of all vectors to their cluster's mean is minimized. 
In the third and final step, the classification result was visualized by marking pixels the spectra of which belong to the same cluster with the same color, yielding a false-color image that very descriptively shows which domain or phase is present at which location on the surface. 

For the low-T sample, Fig.\ \ref{fig:3}(b) displays the result of the clustering with $K=3$. Fig.\ \ref{fig:3}(c) to (e) show the large number of spectra in each cluster as a colored bunch of highly transparent curves. The white lines represent the average of all curves in each of the clusters; they are also plotted in Fig.\ \ref{fig:3}(f) in the color of their cluster. 
Overall, we find that all spectra are rather similar, even across the three clusters. With the possible exception of cluster II (see below), the spectra have one minimum and thus reveal that the corresponding regions on the surface are covered by precisely one decoupled graphene layer \cite{Hibino2008, Ohta2008, Luxmi2010, Mende2018, Riedl2009}. A more detailed analysis of the three clusters, to which we now turn, confirms this finding. 

\begin{figure*}
\includegraphics[width=\textwidth]{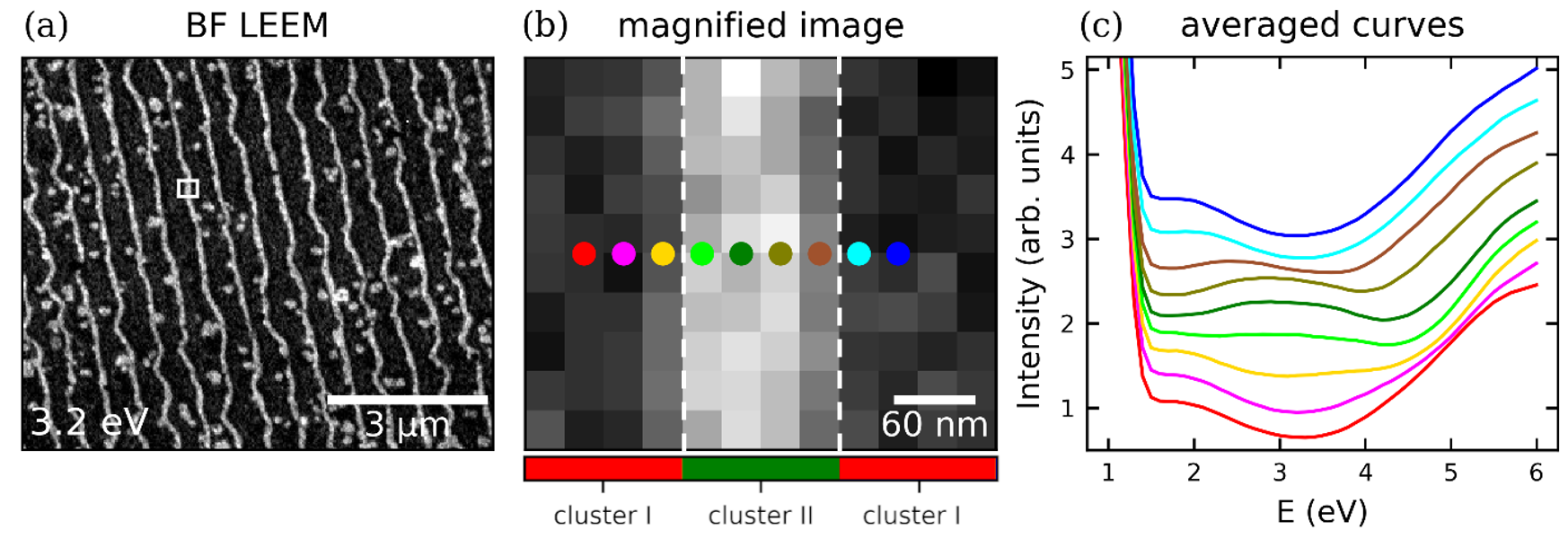}
\caption{\label{fig:4}(a) Bright-field (BF) LEEM image of the same area as shown in Fig.\ \ref{fig:3}(a), but recorded at $3.2$~eV; at this start energy, the contrast between step edges and terraces is more pronounced. (b) Magnified part of the image in panel (a), marked by a white rectangle in (a), showing a vertical step edge. The image size is $360 \times 300$~nm$^2$ ($12\times10$ pixels). The colored dots mark nine columns (each containing 10 pixels) oriented parallel to the step edge. Red and green bars at the bottom of the image indicate to which cluster the pixels in the corresponding column have been assigned in the \textit{K-means} analysis; vertical dashed white lines mark the borders between cluster I (red) and II (green). (c) Averaged LEEM-$IV$ intensities from the marked columns in panel (b). Neighboring spectra are vertically displaced by $0.3$. The color coding corresponds to the colored dots in panel (b).}
\end{figure*}

The characteristic features of the curves in cluster I are a shoulder between $\sim 1.5$~eV and $2.0$~eV and a relatively deep minimum at $\sim 3.2$~eV. As mentioned above, these are the typical features of monolayer graphene on SiC \cite{Hibino2008, Virojanadara2008, Luxmi2010, Riedl2009}. The false color image in Fig.\ \ref{fig:3}(b) indicates that cluster I represents the dominant structure on the surface that covers the terraces almost completely (red). In contrast to clusters I and III, the averaged spectrum of cluster II is essentially flat in the wide energy range from $\sim 1.5$~eV and $\sim 4.0$~eV, although the individual curves show various features---and even maxima and minima---in this region. This cluster is found along the step edges on the surface (green), and at some locations on the terraces where a bright defect is visible in the BF-LEEM image in Fig.\ \ref{fig:3}(a). The origin of this cluster is discussed in more detail in the next paragraph. The spectra in cluster III appear to be intermediate between those of clusters I and II: There is a clear minimum at $\sim 3.2$~eV (like cluster I), but below $2$~eV and above $4$~eV, its average spectrum coincides with the one of cluster II (Fig.\ \ref{fig:3}(f)).
The corresponding areas (blue) on the sample surface are small and match with the dark defects visible in the BF image in Fig.\ \ref{fig:3}(a). Note that at $E\simeq2.5$~eV, marked by the black dashed line in Fig.\ \ref{fig:3}(f), a clear contrast between all three clusters is visible. This is not the case for almost all other energies below $6$~eV. 

We now analyze cluster II in more detail. To this end, we trace the spectra and their changes across a step edge. Figure \ref{fig:4}(a) displays the same surface area as Fig.\ \ref{fig:3}(a), but recorded at an electron energy of $3.2$~eV to obtain a better contrast. Figure \ref{fig:4}(b) shows a small section of the surface, marked in Fig.\ \ref{fig:4}(a) with a white rectangle. It is so strongly magnified that individual pixels appear as squares. A step edge is running through the image from top to bottom. 
In this area, we analyzed the change of the spectra across the step edge by averaging the pixels in the image column-wise, i.e., by summing up the intensities of 10 pixels vertically. Performing this for the nine columns which are marked by colored dots in Fig.\ \ref{fig:4}(b), we obtained the nine spectra that are displayed in Fig.\ \ref{fig:4}(c). We observe a continuous change of the spectra across the step edge, from red via magenta, yellow, different shades of green, brown to cyan and blue. The first and last, red and blue, are almost identical and resemble the spectra of cluster I. In contrast, the dark green, olive and brown spectra clearly show \textit{two} minima, one below 2\,eV, the other shifting from $\sim 4.3$\,eV to $\sim 3.8$\,eV in the sequence from dark green to brown. Remarkably, these are the energies at which surfaces with \textit{two} decoupled graphene layers exhibit their LEEM-$IV$ minima (cf.~Fig.~\ref{fig:9}(a) below). It is clear that the superposition of spectra with one minimum at $\sim 3.2$~eV and two minima, one to left and the other to the right, will yield the flat average spectrum that is characteristic of cluster II.      

To put this finding into perspective, we note that the resolution of our microscope is better than $10$~nm, while the size of one pixel in Fig.\ \ref{fig:4}(b) is $30\times30$~nm$^2$. Hence, our measurement is not resolution-limited in the sense that a (hypothetical) abrupt change of the $IV$ curve at the step edge would be smeared out by the instrumental resolution. In other words, if the change in the spectra at the step edge was abrupt, we would see an immediate change from one column of pixels to the next; the step would appear to be only one pixel wide in the measurement. Evidently, this is not the case, we rather see a continuous change of the spectra in $4$ or $5$ neighboring pixel columns, i.e., on a length scale of approx.\ $120$~nm to $150$~nm. This implies that the continuous, steady change of the curve profile in Fig.\ \ref{fig:4}(c) must be real. 

In light of the above resolution argument, we conclude that the apparent two-layer signal in the step region in Fig.\ \ref{fig:4}~(c) indeed indicates the local formation of bilayer graphene. Thus, a second decoupled graphene layer is formed near the SiC step edges. Considering that for the formation of graphene Si atoms have to evaporate from the bulk surface of SiC, it is plausible that this process starts directly at the step edge, on the \textit{upper} terrace (but not on the lower terrace). Since the steps on the surface of a 6H-SiC(0001) crystal are rather high (three SiC bilayers, corresponding to half the unit cell, i.e. $7.55$~\AA, see above), evaporation of Si at step edges should indeed be favored over evaporation from flat areas on the terraces. The lateral length scale of this effect is of the order of $150$~nm, indicating a moderate diffusion length for the Si surface species, assuming that they desorb only at the edge of the upper terrace.  
In this context, it is noteworthy that the two-layer signal in Fig.\ \ref{fig:4}~(b) appears slightly off-center; based on the model above, this must be toward the upper terrace. Although it is hard to say whether the \Gnull layer overgrows the substrate steps as a continuous layer that bends over the step edge or not, the DF-LEEM results shown above (image \textit{a} in Fig.\ \ref{fig:2}) clearly indicate that there is no (lateral) gap in the \Gnull layer, since the substrate steps are only visible as faint dark lines. Thus, the \Gnull layer obviously extends right to the step edge on both the upper and lower terraces, and thus, looking at the surface from above (along the incident electron beam), no gap is detectable in the \Gnull layer. However, in the DF-LEEM image based on the \Gdrei reflections (image \textit{c} in Fig.\ \ref{fig:2}), bright stripes are visible along the step edges, significantly wider than the faint lines in image \textit{a}, indicating that the \Gdrei structure forms near the step edge (and only there) \textit{in addition} to the already existing \Gnull layer. 
Thus, what we observe at the step edges is the beginning of the decoupling of the \ZLGdrei layer, which transforms into a \Gdrei layer below the \Gnull due to the formation of a new ZLG at the interface with the SiC substrate.

Finally, we discuss the nature of the two types of defects visible in the BF-LEEM image in Fig.~\ref{fig:3}(a). The bright defects have spectra that belong to the same cluster as those recorded at the step edges (cluster II, green). 
It is therefore very likely that we see local tBLG formation also at these defects, possibly induced by some contamination or structural defect in the substrate underneath the graphene layer. The dark defects were assigned to cluster III based on the shape of their LEEM-$IV$ curves, which is somewhat between those of the typical \Gnull (cluster I) and tBLG (cluster II) shapes. However, since the defect contrast in the BF image (Fig.\ \ref{fig:3}(a)) is based on only one specific electron energy, and since clusters II and III in general have rather similar LEEM-$IV$ curve shapes, the differentiation of the two defect types may be exaggerated and to some extent arbitrary. 

We conclude that the surface of the low-T sample is almost homogeneously covered by a uniform \textit{mono}layer of \Gnull (the fact that it is \Gnull rather than \Gdrei follows from DF-LEEM, see section \ref{sec:homogeneity}), in particular on all terraces. Along the step edges, and possibly also at some defects, we observe the beginning growth of $30^\circ$-tBLG, indicated by DF-LEEM on the \Gdrei reflections and the two-layer signature of the $IV$ spectra. The length scale around the steps on which these effects can be observed, is $\sim 150$~nm.

\subsection{Morphology of the high-T \Gnull sample}

\begin{figure}
\includegraphics[width=\columnwidth]{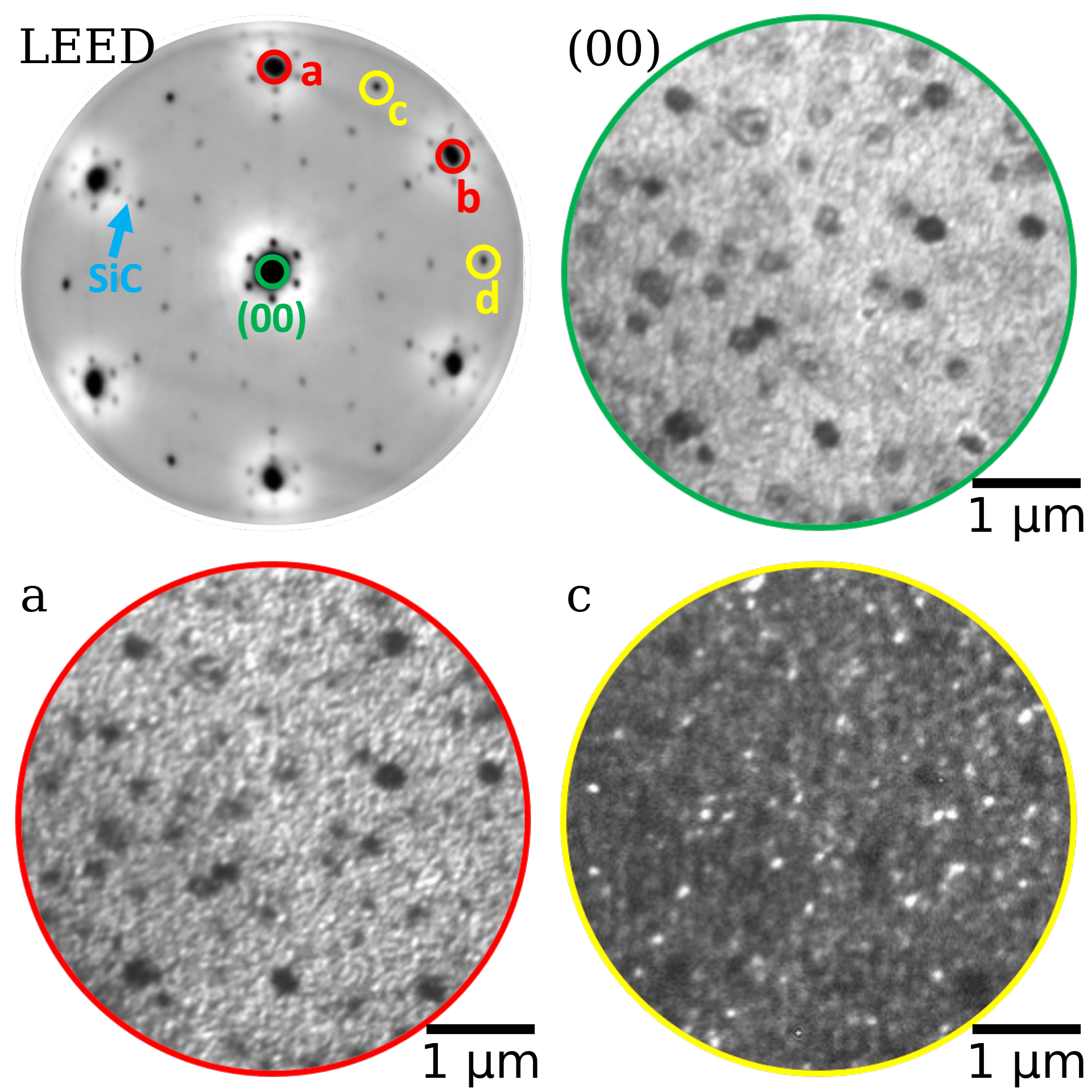}
\caption{\label{fig:5}
LEED pattern of the high-T sample, and corresponding bright-field (BF) LEEM image, labeled  \textit{(00)}, and dark field (DF) LEEM images, labeled \textit{a} and \textit{c}. The labels of the LEEM images correspond to the diffraction spots marked in the LEED pattern. All LEEM images were obtained at an electron energy (start energy) of $E=55$~eV.}
\end{figure}

We now turn to the discussion of the high-T sample, annealed at  $1380^{\circ}$C in borazine atmosphere. We found that the increase of the annealing temperature by $50^{\circ}$C changes the morphology of the resultant sample completely. We observed an inhomogeneous growth of multilayer graphene, as the analysis to which we now turn revealed. 

As for the low-T sample, we start the discussion with LEED, BF- and DF-LEEM results. In the LEED pattern shown in Fig.\ \ref{fig:5}, diffraction spots for both \Gnull and \Gdrei (or \ZLGdrei) are visible and marked by red and yellow circles, respectively. Around the (00) spot (green circle), six satellite spots are visible, stemming from \sixsqthree buffer layer reconstruction. Moreover, two additional rings of diffraction spots are present, each consisting of 12 spots. A similar LEED pattern was reported by Ahn \etal \cite{Ahn2018}, who discussed these spots as a 12-fold symmetric pattern originating from the quasi-crystalline properties of tBLG with 30$^{\circ}$ rotation between the two graphene sheets. Finally, the intensity of the substrate diffraction spots (one position is marked by a cyan arrow) in Fig.\ \ref{fig:5} is very small, too faint in fact to record DF-LEEM images on these spots. This is a first indication that a thicker multilayer stack is present on this sample surface. 

Based on the graphene spots, DF images were recorded at an electron energy of $55$~eV (labeled \textit{a} and \textit{c} in Fig.\ \ref{fig:5}). We also measured a BF-LEEM image, labeled \textit{(00)}. These three images reveal a much higher inhomogeneity of the high-T sample as compared to the low-T sample. Objects of different brightnesses appear to be randomly distributed on the surface (note the different magnifications when comparing the LEEM images in Fig.\ \ref{fig:2} and \ref{fig:5}.)  
Since the DF images based on spots \textit{a} and \textit{b} (as well as \textit{c} and \textit{d}) are  very similar to each other, we show only the images recorded with spots \textit{a} and \textit{c}.  Furthermore, the similarity between images \textit{(00)} and \textit{a} is remarkable, indicating that the \Gnull layer is present everywhere on the sample (similar to the low-T sample). The fact that image \textit{a} is much brighter than image \textit{c} indicates that the \Gnull is the top layer, located above the \Gdrei and/or \ZLGdrei layers (which are also present on the surface, see the LEED image in Fig.\ \ref{fig:5}), with the exception of a few locations on the surface that appear as bright spots in image \textit{c}. Notably, however, the intensity everywhere in image  \textit{c} is non-zero, from which we can conclude that \Gdrei and/or \ZLGdrei must be present on the entire sample surface. 

\begin{figure}
\includegraphics[width=\columnwidth]{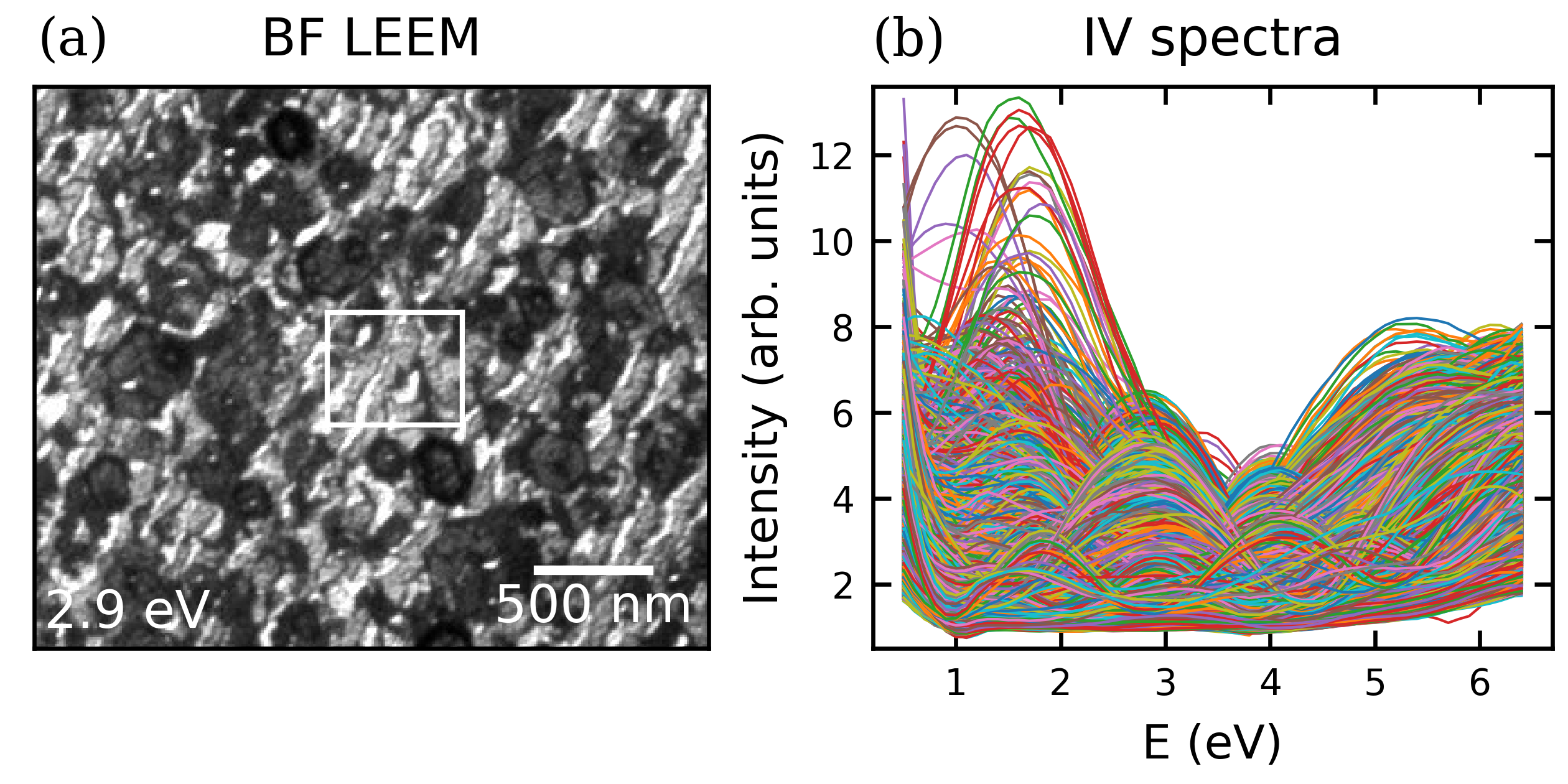}
\caption{\label{fig:6}(a) Bright-field (BF) LEEM image of the high-T sample ($\approx 3.0 \times 2.5$~$\mu$m$^2$, $300\times250$ pixels, $E=2.9$~eV). The white rectangle marks the area that is magnified in Fig.\ \ref{fig:9}. (b) All LEEM-$IV$ spectra that were recorded on the surface area imaged in panel (a). In Fig.\ \ref{fig:7}, these LEEM-$IV$ spectra are clustered. The color coding in panel (b) is arbitrary. }
\end{figure}

\begin{figure*}
\includegraphics[width=\textwidth]{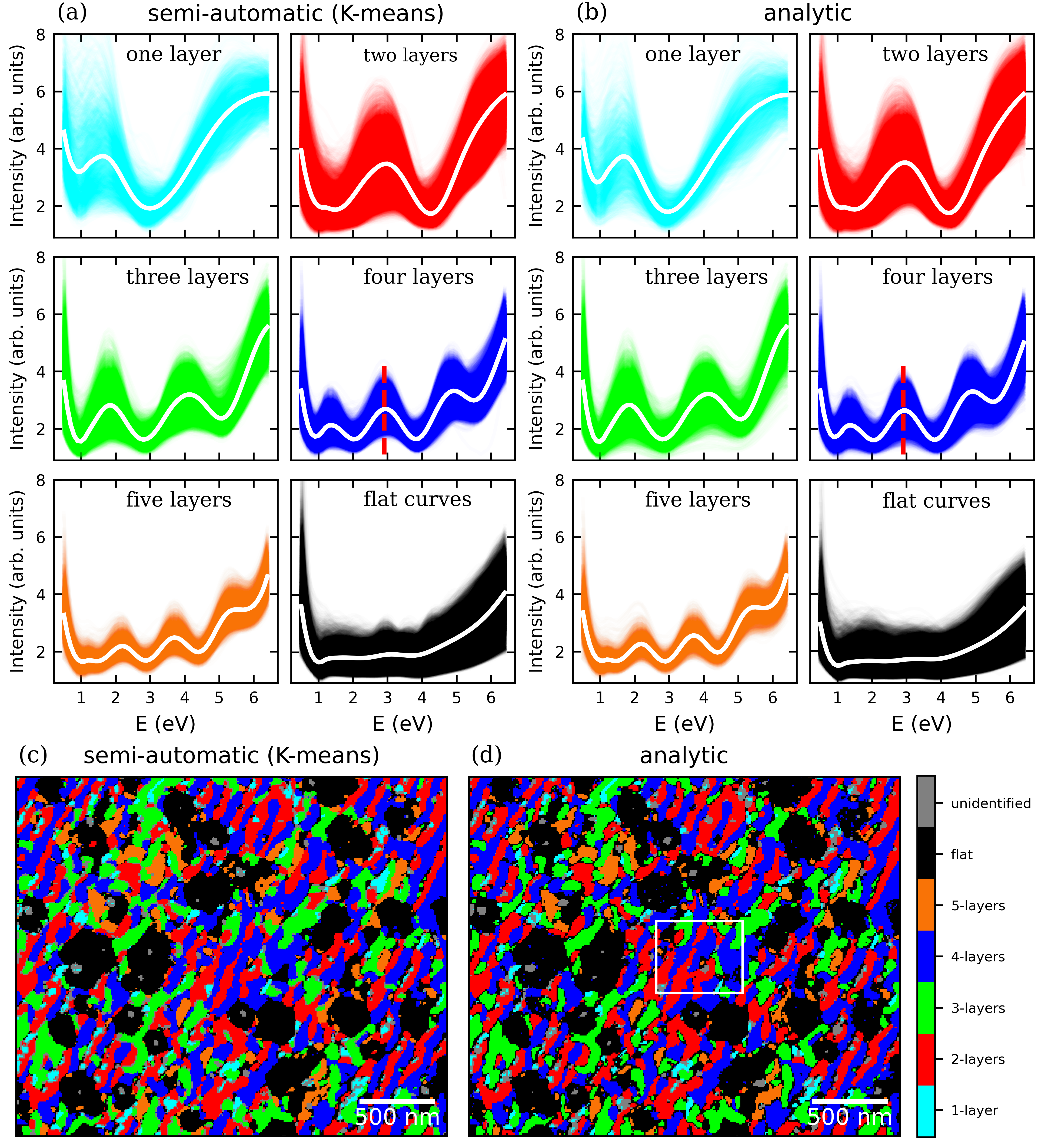}
\caption{\label{fig:7} Results of the clustering of LEEM-$IV$ spectra recorded on the high-T sample for (a,c) the semi-automatic analysis based on the \textit{K-means} algorithm, and (b,d) an analytic classification based on \textit{a priori} defined criteria. For more details, see main text. (a,b) Clusters of spectra for six different classes, originating from graphene stacks consisting of one, two, three, four, five and more than five layers. Individual spectra are plotted with a transparency of $0.01$. The white curves represent the average of all spectra in the respective cluster. (c,d) False-color images representing the result of clustering the LEEM-$IV$ data. The corresponding bright-field (BF) LEEM image is shown in Fig.\ \ref{fig:6}(a). }
\end{figure*}

The key question posed by the LEEM images of the high-T sample in Fig.\ \ref{fig:5} concerns  the origin of the inhomogeneities that are clearly visible in both BF- and DF-LEEM. In order to answer this question, we performed LEEM-$IV$ measurements and analyzed them in a similar way as for the low-T sample. The image in Fig.\ \ref{fig:6}(a) shows the surface area ($300\times250$ pixels) from which the $IV$ spectra were extracted, as well as all $75.000$ spectra plotted in one diagram.
From Fig.\ \ref{fig:6}(b) it is immediately obvious that the shapes of the spectra are very diverse, in particular regarding the number of minima appearing in the relevant energy range.  This indicates that there are regions present on the surface which are covered by different numbers of graphene layers. In the \textit{K-means} analysis that follows, we therefore place great emphasis on identifying the number of minima in the spectra. 

In principle, the \textit{K-means} algorithm is well suited for classifying  curves with complex shapes. However, since the algorithm defines shape similarity by a distance in Euclidean space computed from intensity values (see discussion above), curves will be clustered based on commonality in general features, e.g., (overall) similar intensity, but not on specific attributes such as the number of minima between $E=0$~eV and $6$~eV (which is the only relevant criterion in our case). Therefore, clustering with moderately small $K$ (e.g., $K=15$) failed in our case, because too many spectra with small Euclidean distances but different numbers of minima were grouped together. Performing the classification with a much larger $K$ was more successful. For example, with $K=100$, the variance of the spectra within each cluster became sufficiently small that the number of minima formed a decisive criterion for the Euclidean distance. Obviously, after the initial clustering using \textit{K-means}, a careful (manual) analysis of the results was necessary to group together those clusters that had the same number of minima. Since these are only $100$ clusters instead of $75.000$ spectra, this manual step was a reasonable task. 
The final result of this automated classification with subsequent manual refinement is displayed in Fig.\ \ref{fig:7}(a). Six classes have been identified, five of them allowing an unambiguous counting of minima; a sixth class consisted of flat curves. The false-color image of the relevant surface area shown in Fig.\ \ref{fig:7}(c) displays the results. Note that a small fraction of the curves (6.0\%) did not fit any of these criteria and remained unidentified, although they in turn can be divided into four well-defined sub-classes with distinct $IV$ spectra, see supplementary Fig.\ \ref{fig:S10}. 

Before turning to a detailed discussion of surface morphology of the high-T sample based on the false-color map in Fig.\ \ref{fig:7}(c), we consider an alternative to the semi-automatic, \textit{K-means} based classification discussed above. Considering the large number of spectra, a purely manual classification according to the number of minima is clearly out of question. However, a classification on the basis of a small number of \textit{a priori} defined criteria turned out to be successful. In the present case, these criteria are the number and the positions of the minima in the spectrum. For instance, for the decision whether a given spectrum should be assigned to the two-layers class, the code looked for (exactly) two minima, one located between $1.2$ and $1.5$~eV, the other between $4.0$ and $4.3$~eV. Evidently, this analytic mode of classifications is not prone to false assignments due to a diffuse similarity between spectra, because it relies exclusively on the relevant criteria which are clearly defined  from the outset. 
On the other hand, the semi-automatic classification has the advantage of being model-free, but in the present case it needs a grossly exaggerated number of clusters to start with, followed by a refinement step, in which the clusters obtained by \textit{K-means} are combined into physically meaningful classes. In the analytic classification scheme, this physical meaning is introduced \textit{a priori} by defining physically motivated criteria. In terms of efficiency, the \textit{K-means}-based classification turned out to be  superior in the present case.

\begin{figure}
\includegraphics[width=\columnwidth]{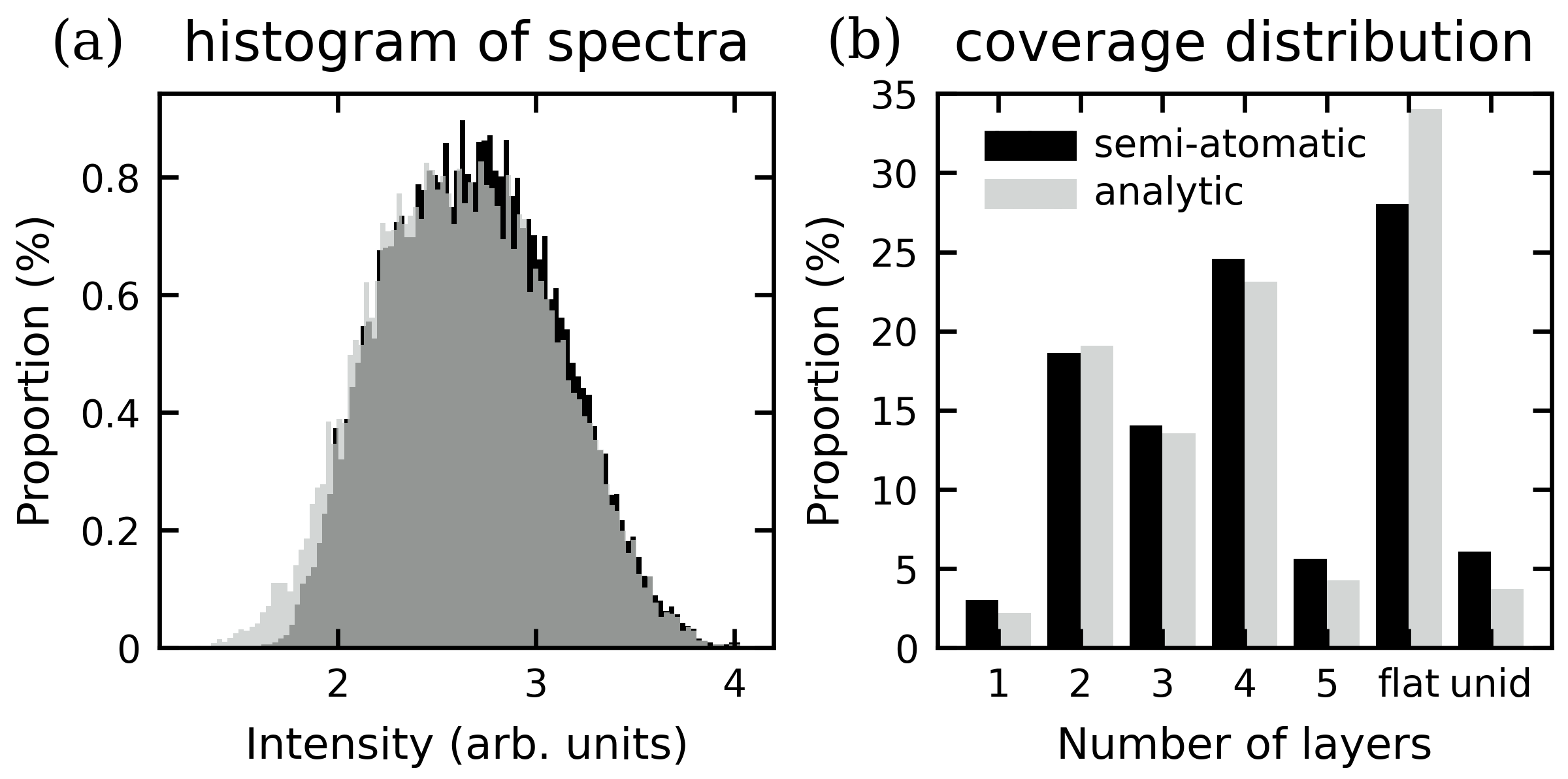}
\caption{\label{fig:8} Results of the clustering of LEEM-$IV$ spectra of the high-T sample for the semi-automatic (black) and the analytic method of data analysis (gray).
(a) Histogram of the LEEM intensity of the four-layer clusters at a start energy of $2.9$~eV, as marked by red dashed lines in Fig.\ \ref{fig:7}(a) and (b). The bin size on the intensity axis was $0.03$. The histogram was normalized to a total area of 1. In the dark gray area the histograms of both  methods overlap. (b) Proportion of the different regions ($1, 2, ..., 5$ layers, more than $5$ layers (flat) and unidentified) of the total surface area, according to Fig.\ \ref{fig:7}(c) and (d). }
\end{figure}

The results achieved by the analytic classification method are shown in Fig.\ \ref{fig:7}(b) and (d). Comparing them with the results of the semi-automatic approach, we observe an excellent agreement between the two, both in the clusters of spectra (Figs.\ \ref{fig:7}(a) and (b)) and in the false-color images in Fig.\ \ref{fig:7}(c) and (d). Only marginal differences appear. For example, in the clusters of spectra belonging to the one-layer and flat-curves classes, the variance varies slightly between Figs.\ \ref{fig:7}(a) and (b). However, at most energies, histograms across the clusters of spectra in each class are unimodal symmetric, which proves an appropriate classification of the spectra in both methods. This is illustrated in Fig.\ \ref{fig:8} for the four-layers class at $E=2.9$~eV (red dashed line in Fig.\ \ref{fig:7}(b)).

\begin{figure*}
\includegraphics[width=\textwidth]{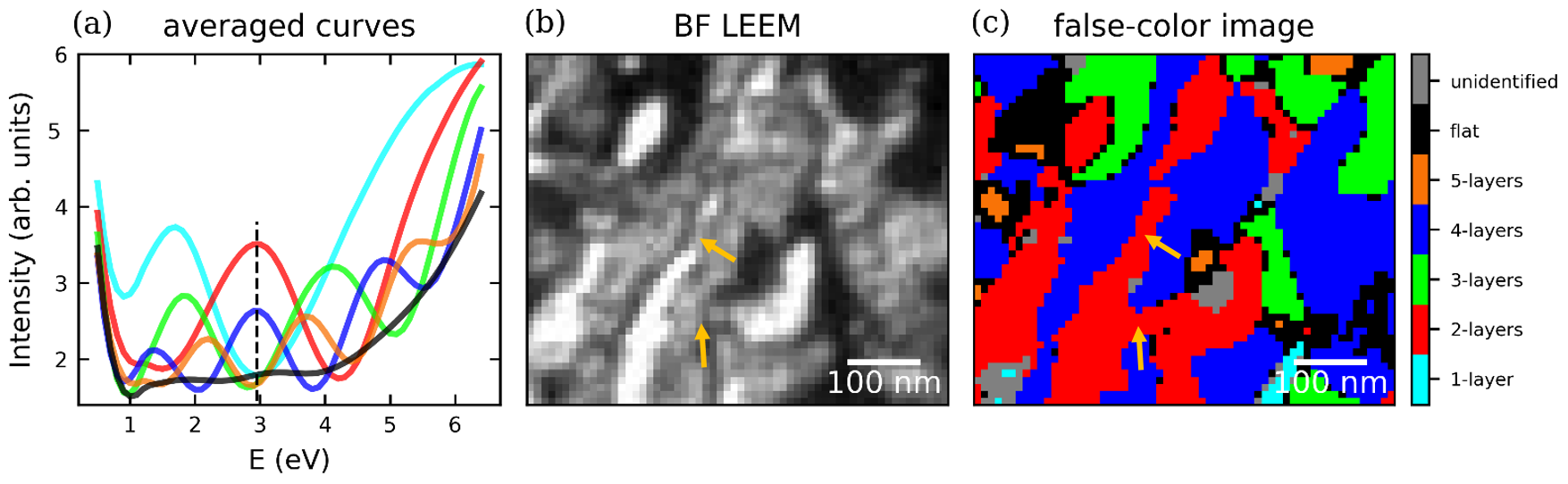}
\caption{\label{fig:9}Results of the clustering (using the analytic approach) of LEEM-$IV$ spectra recorded on the high-T sample: (a) Averaged spectra for each of the classes in Fig.\ \ref{fig:7}(b). (b) Magnification of the marked section (white rectangle) in the bright field (BF) LEEM image in Fig.\ \ref{fig:6}(a). (c) False-color image representing the result of the clustering of LEEM-$IV$ spectra for the sample area shown in (b). }
\end{figure*}

As a cross check, we compare the LEEM-$IV$ false-color maps with BF-LEEM images.
In Fig.\ \ref{fig:9}(a) the averaged $IV$ spectra of the six classes from Fig.\ \ref{fig:7}(b) are plotted in a single graph. The data were taken from the analytic approach, but the curves would not differ significantly if they had been drawn from the semi-automatic analysis. The plot once again reveals that the amplitudes of the intensity oscillations decrease with increasing number of layers. More importantly, it also explains the contrast in the BF-LEEM images in Figs.\ \ref{fig:6}(a) and \ref{fig:9}(b), the latter being a magnified section of the former. At the  energy at which the BF image was recorded ($E=2.9$~eV, black dashed line in Fig.\ \ref{fig:9}(a)), the two- and four-layer spectra have maxima, the one of the two-layer spectrum being more intense. All other curves exhibit  a minimum at this energy. It is therefore clear that in the BF image the two-layer regions should appear brightest, while those with four layers should exhibit intermediate brightness and all other regions should be dark. This is precisely what is observed if we compare the false-color image from LEEM-$IV$ (Fig.\ \ref{fig:9}(c)) with the BF-LEEM image (Fig.\ \ref{fig:9}(b)): 

The correspondence is almost one-to-one for these three brightness levels. Only in a few small regions (two are marked by orange arrows), mostly located between regions for which the assignment is unambiguous, the LEEM-$IV$ analysis reveals the number of layers more clearly than the BF-LEEM contrast. This is of course due to the fact that the LEEM-$IV$ analysis is based on a full three-dimensional data stack $I_{x,y}(E)$, which as such is able to yield more information than a single BF-LEEM image at one specific energy, even if this energy is well selected in terms of maximizing the contrast between different multilayer regions. We can thus conclude that Fig.\ \ref{fig:9} reveals overall consistency between BF-LEEM images and the clustering analysis of LEEM-$IV$ data.

Finally, we summarize the surface morphology of the high-T sample. Figure \ref{fig:8}(b) shows the distribution of coverage with graphene layers. The surface is mostly covered by regions with two, three, and four layers of graphene ($\approx 55$\%), whereby the two-layer regions, with a coverage of almost $20$\%, represent the desired tBLG (red in \ref{fig:7}(c) and (d)). Note that the large regions having flat curves (black in \ref{fig:7}(c) and (d)) include not only domains with six and more layers, but also all defective regions on the surface.

\section{Conclusion}

In this paper, we have studied the growth of graphene on SiC using the earlier-reported surfactant-mediated growth method of annealing in borazine atmosphere \cite{Bocquet2020}. A central motivation of this work was the question whether twisted bilayer graphene (tBLG) can be grown on SiC by thermal annealing in borazine atmosphere at slightly elevated temperature in comparison to the one used for growing single graphene layers in the unusual, $0^\circ$-rotated orientation on SiC(0001) (i.e., \Gnull).

In agreement with earlier work \cite{Bocquet2020}, we found the sample surface covered with a homogeneous \Gnull layer if the annealing in borazine atmosphere took place at $1330^{\circ}$C. While we cannot unambiguously say whether or not the \Gnull layer overgrew the substrate step edges, we see clear indications for a beginning formation of a (decoupled) \Gdrei layer in a $\sim 150$~nm wide stripe along the step edge, most probably on the upper terrace. Confirmed by both the shape of the LEEM-$IV$ spectra and by DF-LEEM we see both \Gnull and \Gdrei in this area. In other words, we observe the start of $30^\circ$-tBLG growth at the step edges, already at a temperature of $1330^{\circ}$C. However, the majority of the surface area is still covered by the high-quality \Gnull/\ZLGdrei stack, with only a low density of point defects. 

The high-quality and flat morphology of the \Gnull/\ZLGdrei stack suggests the possibility to grow tBLG at slightly elevated temperatures at which the \ZLGdrei is known to convert to \Gdrei. We tested this concept by annealing SiC(0001) in borazine atmosphere at $1380^{\circ}$C, i.e., only $50^{\circ}$C higher than the temperature that leads to the high-quality \Gnull/\ZLGdrei stack. This experiment has been partially successful. Because the uppermost layer across the complete high-T sample was still \Gnull, and since all graphene layers that grow below this are of type \Gdrei, the complete high-T sample was indeed covered with a twisted \Gnull/\Gdrei stack. However, the morphology of this sample was far from that of an ideal tBLG stack. Rather, the high-T sample exhibited a patchwork of small regions with different numbers of \Gdrei layers below \Gnull. Up to five graphene layers could be identified, but there were regions on the sample which probably consisted of even more layers. In fact, this complex sample morphology posed a considerable challenge to our LEEM analysis, which we mastered by analyzing LEEM-$IV$ data pixel-by-pixel. This allowed the unambiguous identification of the number of layers on the surface, with a lateral resolution of $\sim 10$~nm, corresponding to a single pixel in the current setup. 

We thus found that while borazine surfactant-mediated growth of \Gnull graphene monolayers is a self-limiting process, this is not true for the growth of additional \Gdrei layers below the \Gnull. Even at $1330^{\circ}$C we find traces of initial \Gdrei growth at the step edges. At slightly higher temperatures ($1380^{\circ}$C), \Gdrei multilayers grow in an uncontrolled manner, producing a very complex morphology. This growth of \Gdrei is not only influenced by temperature: In additional experiments at similar temperatures and longer annealing times, we saw that the multilayer areas grew with time. In particular, in all cases multilayer regions with more than two layers started to grow before the bilayer regions had spread over the entire terraces. Therefore, we have to conclude that thermal growth of high quality tBLG does not seem to be possible, since the multilayer growth (with more than two graphene layers) starts before a homogeneous tBLG sample can be obtained. As a consequence, we propose that decoupling the $30^\circ$-rotated ZLG by intercalation with a suitable atomic species may be a better strategy to obtain tBLG on SiC(0001) than simple annealing in a borazine atmosphere.

%

\begin{acknowledgments}
H.Y., M.H., F.S.T, F.C.B. and C.K. acknowledge funding by the DFG through the SFB 1083 Structure and Dynamics of Internal Interfaces (project A12).
\end{acknowledgments}

\section*{Author contributions}

F.S.T., F.C.B., and C.K.\ conceived and designed the research. M.H.\ prepared the sample, carried out the ARPES measurements and analyzed the ARPES data. H.Y.\ performed the LEEM experiments. H.Y.\ and F.C.B.\ analyzed the LEEM data with input regarding the clustering from C.W. All authors discussed the results. H.Y., M.H., F.S.T., F.C.B., and C.K.\ conceptualized the results and wrote the paper.

\appendix
\onecolumngrid
\newpage
\section*{Supplementary Figure}

\begin{figure}[h!]
\includegraphics[width=\textwidth]{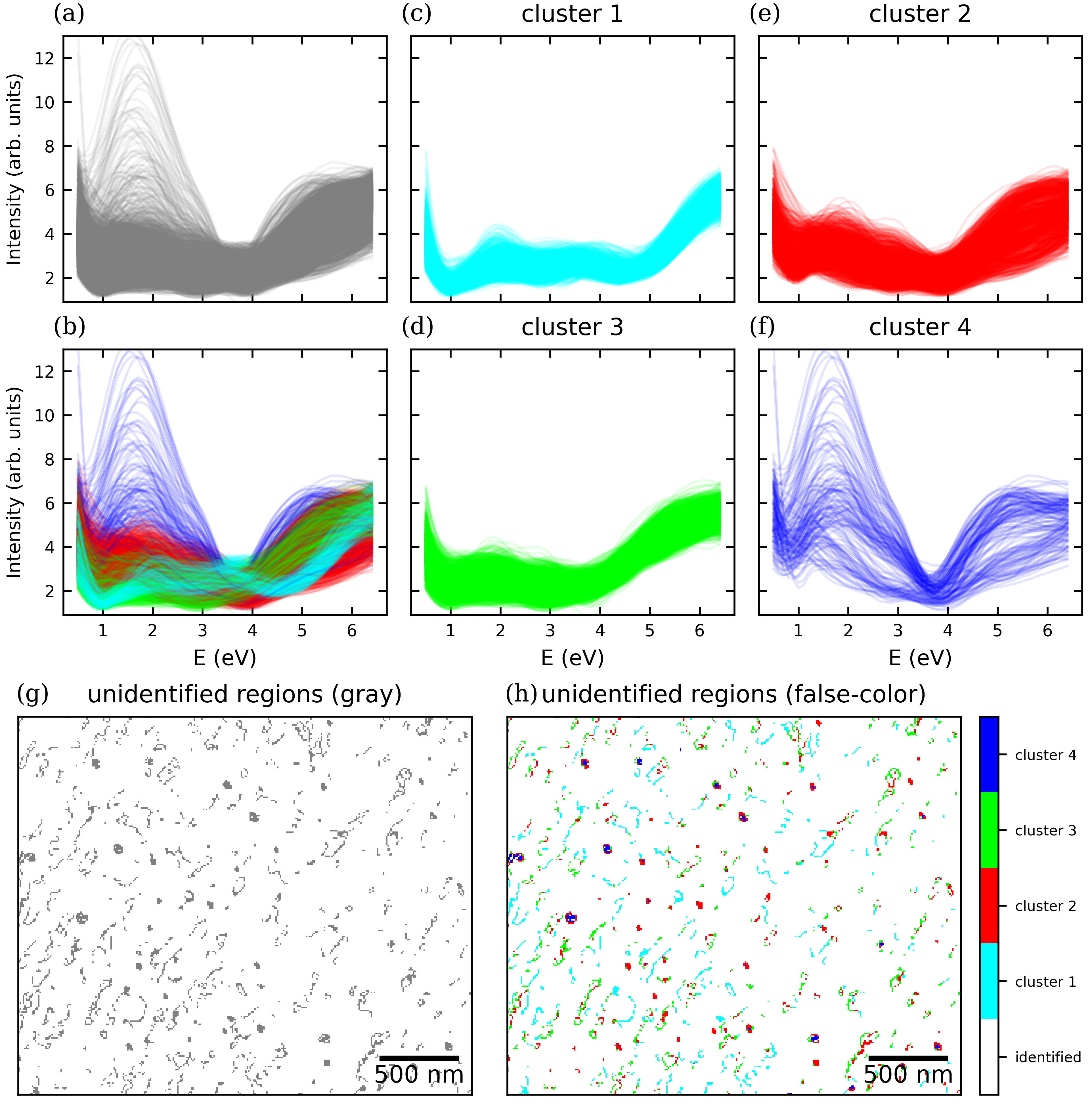}

\caption{\label{fig:S10} LEEM-$IV$ spectra and false-color images of the unidentified regions found in the semi-automatic (\textit{K-means}) analysis of the high-T sample (compare with Fig.\ \ref{fig:7}). (a) All $IV$ curves of the unidentified regions, plotted in gray with a transparency of $0.01$. The unidentified regions can be divided in four clusters, as shown in (c-f), according to the shape of their $IV$ curves. (b) Same as (a), but all curves plotted in their respective false-color as defined in (c-f). (g) and (h): Areas on the surface corresponding to unidentified regions, plotted as gray and false-colored areas, respectively. The shown surface area corresponds to Fig.\ \ref{fig:7}(c). }
\end{figure}

\end{document}